\documentclass[a4paper, 10pt, twocolumn]{article}

\usepackage[utf8]{inputenc}
\usepackage{amsfonts,amssymb,amsmath,amsthm}
\usepackage{graphicx}
\usepackage[footnotesize]{caption}
\usepackage{siunitx}
\usepackage{subfig}

\newcommand{\vect}[1]{\boldsymbol{#1}}

\newcommand{\xb}{\vect{x}}

\newcommand{\yb}{\vect{y}}
\newcommand{\alphab}{\vect{\alpha}}

\newcommand{\mat}[1]{\boldsymbol{#1}}

\newcommand{\Hb}{\mat{H}}

\newtheorem{proposition}{Proposition}

\newcolumntype{M}[1]{>{\centering\arraybackslash}m{#1}}

\usepackage[top=1.5cm, left=1.5cm, right=1.5cm, bottom=1.5cm]{geometry}

\title{Local-mean preserving post-processing step for non-negativity enforcement in PET imaging: application to \textsuperscript{90}Y-PET.}

\author{Maël~Millardet$^{1,2}$, Saïd~Moussaoui$^1$, Diana~Mateus$^1$, Jérôme~Idier$^1$ and Thomas~Carlier$^{2,3}$.\\
\footnotesize $^1$LS2N, CNRS UMR 6004, Nantes, France.\ $^2$CRCINA, INSERM UMR 1232, Nantes, France.\
$^3$Nuclear Medicine Depart., Univ. Hospital, Nantes, France.
}
\date{\empty} 

\renewenvironment{abstract}{\bf\small {\em\ Abstract---}}{}

\begin{document}

\maketitle

\begin{abstract}
In a low-statistics PET imaging context, the positive bias in regions of low activity is a burning issue. To overcome this problem, algorithms without the built-in non-negativity constraint may be used. They allow negative voxels in the image to reduce, or even to cancel the bias. However, such algorithms increase the variance and are difficult to interpret, since negative radioactive concentrations have no physical meaning. Here, we propose a post-processing strategy to remove negative intensities while preserving the local mean activities. Our idea is to transfer the negative intensities to neighboring voxels, so that the mean of the image is preserved. The proposed post-processing algorithm solves a linear programming problem with a specific symmetric structure, and the solution can be computed in a very efficient way. Acquired data from an yttrium-90 phantom show that on images produced by a non-constrained algorithm, a much lower variance in the cold area is obtained after the post-processing step.
\end{abstract}

\section{Introduction}
\label{sec:introduction}

Positron Emission Tomography (PET) is a widely used imaging technique of nuclear medicine, used in particular for the management of patients in oncology. It is a native quantitative tool very attractive for revealing molecular processes. However, the quantitative precision of PET suffers from several problems when the number of coincidences recorded is low. In particular, micro-spheres of $^{90}$Y are indicated for the treatment of primary and secondary liver cancers, but $^{90}$Y is a $\beta^-$ emitter that emits positrons with a branching ratio of only \num{3.2e-5}  \cite{thomas:selwyn2007}, and $^{90}$Y-PET imaging therefore suffers from quantification artifacts. Among those, low-activity regions (\emph{i.e.}, cold areas) suffer from systematic positive bias. Reconstruction algorithms allowing negative activity values have been proposed to overcome this issue   \cite{mael:byrne1998,mael:nuyts2002_NEGML,mael:vanslambrouck2015,mael:lim2018}. However, such algorithms increase the variance and introduce non-physical negative values. Here, we propose a non-negativity enforcement post-processing step (NNEPPS) to reduce both the bias and the variance while getting a non-negative image.

Let us start by observing that voxel-wise intensities are not really meaningful. It is rather the mean intensity of small homogeneous regions that conveys physical information.
Thus, we propose to transfer negative voxel values to neighboring voxels. A first idea would be to use the formalism of optimal transport  \cite{mael:monge1781,mael:kantorovitch1942}, to ``transport'' the negative activities towards positive voxels with a minimal cost. However, a major drawback of optimal transport, in this case, is that the transfer may be asymmetric, leading to a spurious spread of information. The approach followed here is to pre-define a symmetric \emph{voxel spread function} describing the proportion of the value coming from a voxel to be distributed to each neighbor. The solution is the one corresponding to the minimal transfer following such a symmetric rule.

The contribution of this work is firstly to formalize this post-processing strategy, and secondly to implement and test the method. We cast the problem in a linear programming framework \cite{mael:nocedal1999}, and we propose an adapted version of the dual simplex algorithm to solve the optimization problem.

This document is an abridged version of the published paper~\cite{millardet2020local}, to which the reader is referred for more details, as well as for the source code.

\section{Problem statement}
\label{sec:problem_statement}

The goal of the proposed post-processing step is to obtain a non-negative image by a minimal spread of the negative values over the positive voxels. Every voxel value is allowed to increase (we call this increase the \emph{transfer coefficient} assigned to this voxel) while its neighboring values decrease such that the local mean is preserved.

This can be formalized in the following way:
\begin{equation} \label{eq:formulation}
\underset{\alphab}{\min} \, ||\alphab|| \
\text{ such that }
\left\{
    \begin{array}{l}
        \alphab \geq 0 \\
        \yb = \xb + \Hb \alphab \geq 0 \\
    \end{array}
\right. 
\end{equation}
where $\xb$ represents the initial image and $\yb$ the final one. $\alphab$ represents a map of the transfer coefficients, and $\Hb$ the operator giving the influence of the transfers on the image. The non-negativity of $\alphab$ is imposed because there is no reason for positive voxels to spread a potentially ``too large'' value. Numerical tests have confirmed the benefit of this constraint. An interesting property is that the final image is the same if the norm to be minimized is the L1 norm, the L2 norm, or actually, any norm strictly increasing according to each coordinate. The preservation of the global mean is automatically fulfilled in this formalism because it involves only transfers. \emph{i.e.}, each line of $\Hb$ sums to zero.
Although not mandatory, a simple choice for $\Hb$ is that it results from a spatially invariant elementary mask which is convoluted with the transfer map $\alphab$, for example, $(-0.5, 1, -0.5)$ for a 1D image. We now state four propositions. For detailed proofs, we refer the reader to the published article~\cite{millardet2020local}.

\begin{proposition} \label{theo:existence}
A solution to (\ref{eq:formulation}) exists if and only if the mean value of the initial image is non-negative.
\end{proposition}

\begin{proposition} \label{theo:uniqueness}
If a solution to (\ref{eq:formulation}) exists, it is unique.
\end{proposition}

\begin{figure}[htb]
\centering
\includegraphics[width=\columnwidth]{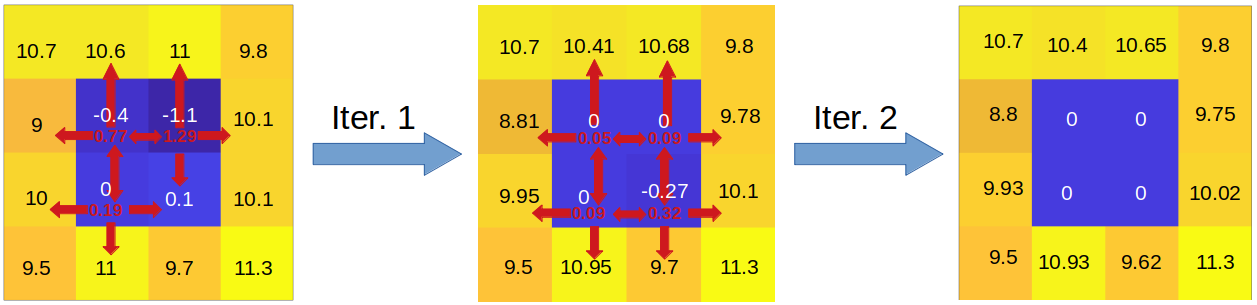}
\caption{In the initial $4\times4$ image (left), three pixels are negative. Their intensities are transferred to their neighbors at iteration 1. A new negative pixel appears (middle). It is added to the set of pixels that need to be subsequently set to 0. After iteration 2 (right), the algorithm stops.}
\label{fig:exemple-2D}
\end{figure}

Fig.~\ref{fig:exemple-2D} illustrates the effect of the NNEPPS on a simple 2D-example.

This problem can be formulated as a linear programming problem. Although 3D-PET images typically contain around \num{e7} voxels, this problem may be solved efficiently by a dual-simplex algorithm, which has, in this case, very interesting properties, and makes the post-processing achievable in around 2 minutes for whole PET images.

\begin{proposition} \label{theo:prop_convergnce_simplex} 
If the post-processing problem has a solution, the dual-simplex algorithm reaches this solution.
\end{proposition}

\begin{proposition} \label{theo:prop_dual_simplex}
Starting from $\yb^{(0)} = \xb$, in every dual-simplex iteration, all the possible entering indexes (all $i$ for which $y^{(k)}_i < 0$) are in the inactive set of the solution \emph{i.e.}, all possible choices are correct choices. On top of that, the leaving index associated to an entering one is known.
\end{proposition}

\section{Results}

\begin{figure}[htb]
\centering
\includegraphics[width=0.46\columnwidth]{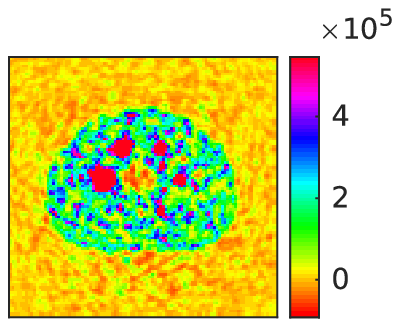}
\quad
\includegraphics[width=0.46\columnwidth]{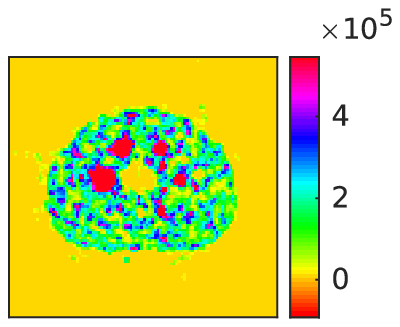}
\caption{Reconstruction of a phantom with AML before and after the NNEPPS (PP-AML, for Post-Processed AML). In PP-AML, the cold area at the center is much clearer.}
\label{fig:image_ss_neg_non_saturee}
\end{figure}

\begin{figure}[htb]
\centering
\includegraphics[width=.93\columnwidth]{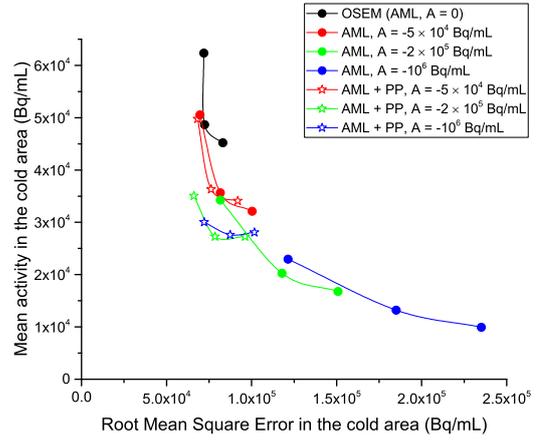}
\caption{Mean activity of the central cylinder (theoretical value: 0) as a function of the RMSE in the same area (theoretical value: 0), for different reconstruction algorithms before and after the NNEPPS. Each point in the same curve represent a different iteration number (from 1 to 3)}
\label{fig:image}
\end{figure}

\begin{figure}[htb]
\centering
\includegraphics[width=.93\columnwidth]{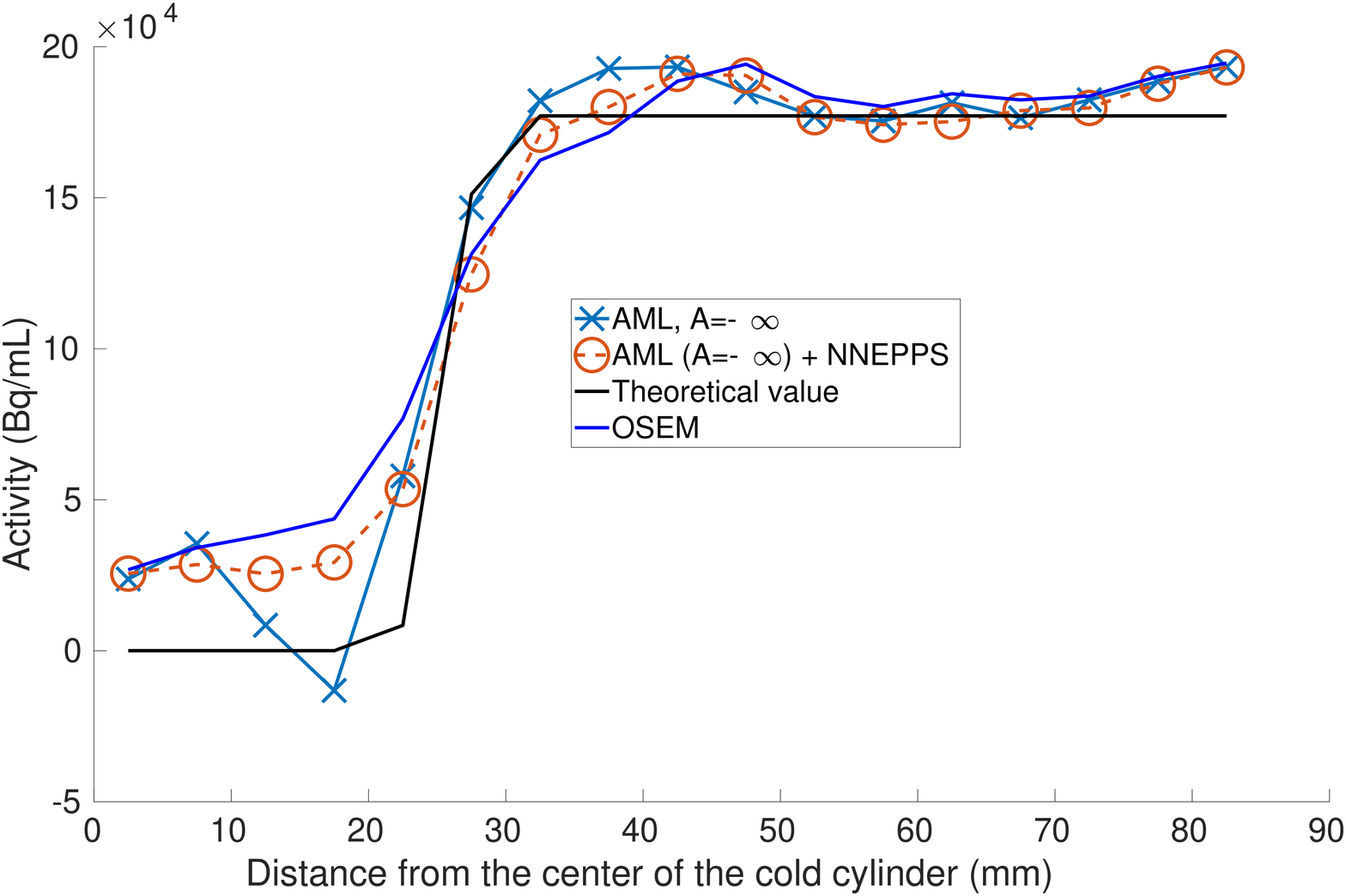}
\caption{Graphs displaying the mean activity measured as a function of the distance from the center of the cold cylinder. The theoretical value and OSEM are given for reference.}
\label{fig:cylindre-autour}
\end{figure}

Fig.~\ref{fig:image_ss_neg_non_saturee} shows the image produced by the unconstrained algorithm AML\footnote{AML stands for Maximum likelihood with a lower bound A. This lower bound can be made negative to allow negative values}~\cite{rahmim2012direct} before and after the NNEPPS. This phantom (NEMA PET IEC) is composed of a warm background (mostly green in the images), in which an activity of \SI{177}{\kilo\becquerel\per\milli\liter} has been introduced. Six spheres (in red) were filled with an activity of \SI{1,33}{\mega\becquerel\per\milli\liter}. A cylinder was also preserved from any activity. Visually, the NNEPPS cleans all the noise present in the areas free from activity, while keeping unchanged other areas.
Figure \ref{fig:image} shows that the RMSE in the cold cylinder is roughly divided by 2. The positive bias remains approximately unchanged by the NNEPPS until it reaches a limit  (of approximately \SI{2.5e4}{\becquerel\per\milli\liter} in this example).
Figure \ref{fig:cylindre-autour} shows that the NNEPPS not only achieves the desired goal but also creates a small negative bias in the part of the warm area in the neighborhood of the cold cylinder. However, this negative bias is very limited and the activity remains higher than the constrained algorithm Ordered Subsets Expectation Maximization (OSEM)~\cite{mael:hudson1994}, which can be taken as a reference due to its popularity and its use in clinical routine.

\section{Conclusion}

In this work, a post-processing step removing negative intensities is introduced. It aims to keep the bias reduction while reaching a low variance. The proposed algorithm is a condensed simplex on the dual problem, which is justified by the particular nature of the problem. Compared to unconstrained algorithms that produce unbiased outcomes, the NNEPPS, by the spread of negative values into/outside a specific region, generally creates a small bias, especially at the boundary between regions. However, this bias remains lower than that of constrained algorithms. In cold areas, the RMSE is highly improved by the NNEPPS. From a qualitative point of view, images are also visually easier to interpret after the NNEPPS. In particular, cold areas are much more visible.

\section*{Acknowledgment}

This work has been supported in part by the European Regional Development Fund, the Pays de la Loire region on the Connect Talent scheme MILCOM (Multi-modal  Imaging  and  Learning for  Computational-based  Medicine),  Nantes M\'etropole (Convention 2017-10470), the French National Agency for Research called "Investissements d'Avenir" IRON Labex n\textsuperscript{o} ANR-11-LABX-0018-01 and the French program Infrastructure d’avenir en Biologie Santé ANR-11-INBS-0006 (France Life Imaging).


{\small
\bibliographystyle{ieeeji}
\bibliography{IEEEabrv,bibliography}
}

\end{document}